\documentclass{iaus}
\usepackage{graphicx}

\title[Interaction of turbulent convection and rotation in RGB stars] 
{On the interactions of turbulent convection and rotation in RGB stars}

\author[A. Palacios and A.S. Brun]   
{Ana Palacios$^1$ 
\and Allan S. Brun$^1$}

\affiliation{$^1$CEA/DSM/DAPNIA/Service d'Astrophysique and AIM UMR 7158, CE Saclay B\^at
  709, F-91191 Gif-sur-Yvette, France \break email: ana.palacios@cea.fr or sacha.brun@cea.fr}

\pubyear{2007}
\volume{239}  
\pagerange{119-224}
\date{?? and in revised form ??}
\jname{Convection in Astrophysics}
\editors{F. Kupka, I.W. Roxburgh \& K.L. Chan, eds.}
\begin{document}

\maketitle

\begin{abstract}
We have performed the first three-dimensional non-linear simulation of the
turbulent convective envelope of a rotating 0.8 ${\rm M}_\odot$ RGB star
using the ASH code.  Adopting a global typical rotation rate of a tenth of
the solar rate, we have analyzed the dynamical properties of the convection
and the transport of angular momentum within the inner 50\% in radius of
the convective envelope. The convective patterns consist of a small
number of large cell, associated with fast flows ($\sim 3000~{\rm m/s}$)
and large temperature fluctuations ($\sim 300~{\rm K}$) in order to carry
outward the large luminosity ($L \sim 400~{\rm L}_{\odot}$) of the star. The
interactions between convection and rotation give rise to a large radial
differential rotation and a meridional circulation possessing one cell per
hemisphere, the flow being poleward in both hemisphere. By analysing the
redistribution of angular momentum, we find that the meridional circulation
transports the angular momentum outward in the radial direction, and
poleward in the latitudinal direction, and that the transport by Reynolds stresses acts in
the opposite direction. From this 3-D simulation, we have derived an average radial rotation profile, that
we will ultimately introduce back into  1-D stellar evolution code.
\end{abstract}

\firstsection 

\section{Astrophysical context}

The standard\footnote{By this we refer to models of non-rotating,
non-magnetic stars, in which convection and atomic diffusion are the only
transport processes considered.} stellar evolution theory predicts that the
surface chemical composition of low-mass stars is modified after the end of
the main sequence (turn-off) by the deepening of the convective envelope
(first dredge-up), and is kept unaltered after the first dredge-up
completion until they reach the horizontal branch.\\ On the other hand,
observations paint a rather different picture, indicating that not less than
98\% of low-mass red giant branch (hereafter RGB) stars exhibit surface
abundance variations of lithium, carbon and nitrogen on the upper part of
the RGB (Charbonnel \& do Nascimento~\cite{CdoN98}). These ``abundance
anomalies'' appear at the {\em bump}, which is the evolutionary point where
the outgoing hydrogen burning shell crosses and erases the chemical (mean
molecular weight) discontinuity left behind by the retreating convective
envelope after the first dredge-up. They are the result of hydrogen nuclear
burning through the CNO cycle, and have been demonstrated to be clear
signatures of internal transport processes (Gratton
\etal~\cite{Gratton00}). The nature of such internal transport processes
has been related to rotation since the pioneering work by Sweigart \&
Mengel (\cite{SM79}) (see also Charbonnel \cite{CC95}), but it is only up
to recently that consistent treatment of rotation-induced mixing by
meridional circulation and shear turbulence was applied to low-mass stellar
evolution on the RGB. Using the best available tools to describe rotational
mixing, in Palacios \etal~(\cite{PCTS06}) we found that meridional
circulation and shear-induced turbulence, as described by the present
formalism, and taken as ``stand alone'', do not drive sufficient mixing in
the radiative zone so as to modify the surface chemical composition of RGB
stars. One of the weaknesses of the formalism we used is to consider
solid-body rotation in the convective zones. The analysis of the rotation
of horizontal branch stars in globular clusters recently led Sills \&
Pinsonneault (\cite{SP00}) to suggest, as Sweigart \& Mengel (\cite{SM79})
did, that in order to reproduce the rotation of HB stars, a large amount of
angular momentum should be retained in the inner part of the stars during
the RGB ascension. This can be achieved if RGB convective envelopes rotate
differentially. After confirming in Palacios \etal~(\cite{PCTS06}) that
assuming conservation of the specific angular momentum ($\Omega(r) \propto
r^{-2}$) favours turbulent mixing below the convective zone, we have
decided to investigate more deeply the interaction of turbulent convection
and rotation in such an extended convective zone. To do so, we turned to 3D
simulations, that can be used as numerical experiments in order to better
understand such dynamical interactions.

\section{Numerical simulation : a brief description}
 We have used the anelastic spherical harmonic (ASH) code in its
 hydrodynamic mode to study in details the interaction of turbulent
 convection and rotation in the inner part of the convective envelope of a
 low-mass RGB star. The reader is referred to Brun \& Toomre (\cite{BT02})
 and Brun \etal~(\cite{BMT04}) for details on the code. Let us here briefly
 summarize the main characteristics of our simulation. We constructed the
 3D simulation by using as initial reference state, a 1D model of a RGB
 star at the {\em bump} with the following characteristics : ${\rm M}_{\rm
 ini} = 0.8~M_\odot$, ${\rm Z} = 2~10^{-4}$, ${\rm L}_* = 425~L_\odot$ and
 ${\rm R}_* = 40~R_\odot$. In order to reduce the density contrast to an
 amplitude that can be numerically handled by ASH, we reduced the
 computational domain to the inner $50\%$ of the convective envelope, in
 the region ${\rm r} \in [0.05~R_\odot; 0.5~ R_\odot]$, where the
 luminosity is constant and $\Delta \rho = 100$ ($\rho \in [10^{-5};
 10^{-3}]~g.cm^{-3}$).  We assume as initial solid-body rotation $\Omega =
 \Omega_\odot / 10 = 2.6~10^{-7} s^{-1}$. We assume rigid stress free
 boundary conditions at the edges of the computational domain, and present
 here first results for a simulation with a Prandtl number $P_r = \nu /
 \kappa = 1$, and a Reynolds number $R_e = \upsilon L / \nu \approx 500$.
 We impose a flux of radiative energy at the base of our domain that is
 extracted at the surface. The simulation has been evolved over 3230 days,
 which represent more than 10 rotations of the system.

\section{Convective patterns in the inner part of the extended convective envelope of a red
              giant star}

\begin{figure}[t]
\begin{center}
\resizebox{\hsize}{!}{\includegraphics{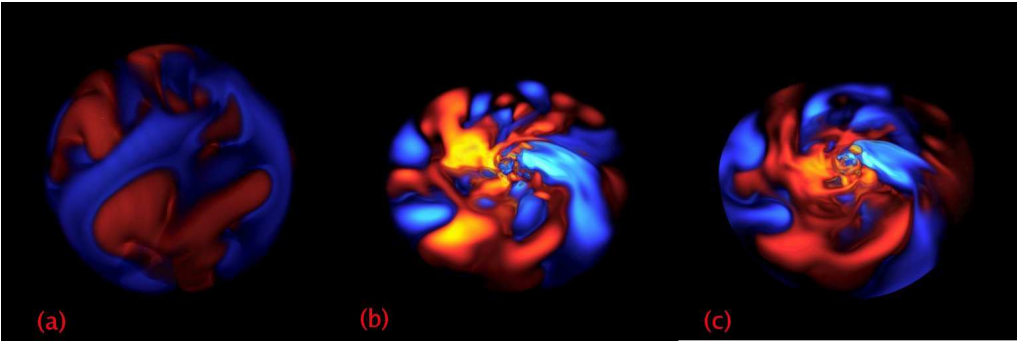}}
\caption{Global volume rendering of the radial velocity flow in a full
  sphere view at the top of the computational domain ($a$), and rendering
  of the radial velocity ($b$) and the temperature ($c$) with the northern
  hemisphere being removed, such as to display the equatorial and inner
  regions. Reddish patterns are warm upward flows and blueish patterns
  are cool downward flows.}
\label{fig1}
\end{center}
\end{figure}

The convective instability sets in rapidly, with a linear growth phase
spanning 130 days after the beginning of the simulation, and non-linearly
saturates, reaching a statistical equilibrium that lasts for the remaining
of the simulated time. The resulting convective pattern is represented in
Fig.~\ref{fig1} : it consists of large blobs covering the top of the
simulation domain, the size of which decreases towards the bottom of the
domain, partly due to a geometrical effect. The downflows are associated
with narrower and cooler structures. There is a clear correlation between
radial velocity and temperature at all depths. The characteristic time for
convective overturning is about 150 days, which is of the order of half the
rotation period. The large luminosity, 400 ${\rm L}_\odot$, associated
with the radiative flux that is applied at the base of the domain is
transported through the spherical shell thanks to the convective (enthalpy)
flux associated with the very large fluctuations of the radial velocity
($\approx 3000~{\rm m/s}$) and of the temperature ($\approx 300~{\rm
K}$).\\ Contrary to simulations under Boussinesq approximation, we obtain
asymmetric convection, with large slow upflows and narrow fast downflows, characteristic of compressible computations. The
enthalpy flux is very large and represents more than 170\% of the imposed
radiative flux (i.e convective luminosity ${\rm L}_{en}$= 4$\pi$ $r^2$
($\bar{\rho}$ Cp $<$ $\upsilon_r$' T$>$) = 1.7 ${\rm L}_*$)in the inner half part of the computational domain. In this
region, it compensates the large negative kinetic energy flux, which can be
there as high as 70\% the imposed radiative flux. This is in sharp contrast with
MLT modelling used in 1-D models, that assumes that the convective
luminosity is equal to the stellar luminosity, and that the kinetic energy
luminosity is zero.

\section{Established differential rotation and angular momentum balance}

\begin{figure}
\begin{center}
\includegraphics[width=0.7\textwidth,clip]{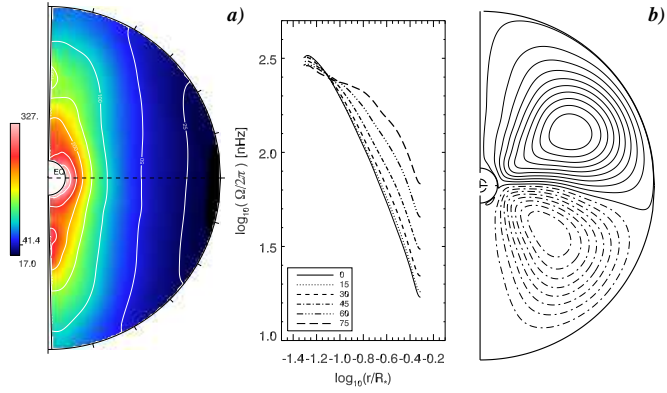}
\caption[]{$(a)$ Temporal and longitudinal averages of the angular velocity
profile achieved in the simulation formed over 1800 days. Radial cuts at
indicated latitudes are presented on a logarithmic scale in the middle
panel. $(b)$ Streamlines of the mean axisymmetric meridional circulation
achieved in our simulation, averaged over 1800 days. Solid contours denote
counterclockwise circulation. The case studied here presents one
poleward cell per hemisphere.}
\label{fig2}
\end{center}
\end{figure}

In this extended convective domain, strong differential rotation develops,
as can be seen from Fig.~\ref{fig2}.a, where we display the temporal (over
1800 days) and longitudinal average of the angular velocity profile
realized in the simulation. Due to the moderate degree of turbulence in the
simulation, the obtained rotation is very cylindrical. A large differential
rotation is achieved both in the radial and latitudinal directions. The
amplitude of the differential rotation in the radial direction is
$\Delta \Omega^r / \Omega_0 = (\Omega_{R_{top}} - \Omega_{R_{bot}}/\Omega_0)
= 7.5$,
the inner prograde regions rotating faster than the outer retrograde
ones. The contrast in angular velocity in latitude is of $(\Delta
\Omega^\theta / \Omega_0)_{top} = (\Omega_{\theta = 0} -
\Omega_{\theta=\pi/2}/\Omega_0) = 1.2$ (where $\theta$ is the colatitude), the high
latitude regions rotating faster than the equator. The rotation that sets in is also anti-solar, with a slow equator
and fast poles.\\

The associated meridional flows are represented in Fig.~\ref{fig2}.b. They
exhibit one large poleward cell at the surface per hemisphere, with a
return flow deeper down.\\
Meridional circulation contributes to a non-negligible part of the total
kinetic energy (KE), which is of the order of a few $10^6~{\rm
erg/cm^3}$. Half of it corresponds to the non-axisymmetric convection
(CKE). The kinetic energy associated with differential rotation (DRKE)
represents 40\% of the total energy while the 10\% remaining are associated
with meridional circulation (MCKE; see Brun \& Toomre \cite{BT02} for their
analytical expression). This energy distribution is very
different from that obtained for simulations of the convective envelope of
the Sun, where MCKE represents less than 1\% of the total kinetic energy,
and exhibits a multi-cellular structure, both in radius and latitude. The
moderate degree of turbulent achieved in this simulation can perhaps
explain the rather simple meridional flow established.\\

In our simulation, the choice of stress-free boundaries at the edges of
the computational domain results in the conservation of the angular
momentum. 
The $\phi$-component of the momentum equation expressed in conservative form and averaged in time
and longitude yields

\begin{equation}
\frac{1}{r^2} \frac{\p(r^2 {\cal F}_r)}{\p r}+\frac{1}{r \sin\theta}
\frac{\p(\sin \theta {\cal F}_{\theta})}{\p
\theta}=0,
\end{equation}
involving the mean radial angular momentum flux
\begin{equation}
{\cal F}_r=\hat{\rho}r\sin\theta[\underbrace{-\nu r\frac{\p}{\p
r}\left(\frac{\hat{v}_{\phi}}{r}\right)}_{{\cal F}_{r,V}}+\underbrace{\widehat{v_{r}^{'}
v_{\phi}^{'}}}_{{\cal F}_{r,R}}+\underbrace{\hat{v}_r(\hat{v}_{\phi}+\Omega_0
    r\sin\theta)}_{{\cal F}_{r,M}}] 
\end{equation}
and the mean latitudinal angular momentum flux
\begin{equation}
{\cal F}_{\theta}=\hat{\rho}r\sin\theta[\underbrace{-\nu
\frac{\sin\theta}{r}\frac{\p}{\p
\theta}\left(\frac{\hat{v}_{\phi}}{\sin\theta}\right)}_{{\cal F}_{\theta,V}}+\underbrace{\widehat
{v_{\theta}^{'} v_{\phi}^{'}}}_{{\cal F}_{\theta,R}}+\underbrace{\hat{v}_{\theta}(\hat{v}_{\phi}+\Omega_0
r\sin\theta)}_{{\cal F}_{\theta,M}}].
\end{equation}

${\cal F}_{r,V}$ and ${\cal F}_{\theta,V}$ are related to the angular momentum
flux due to viscous transport, ${\cal F}_{r,R}$ and ${\cal F}_{\theta,R}$ are
related to the transport through Reynolds stresses and ${\cal F}_{r,M}$ and
${\cal F}_{\theta,M}$ represent the angular momentum flux due to meridional circulation.\\

\begin{figure}
\begin{center}
\includegraphics[height=5.5cm,width=0.5\textwidth,clip]{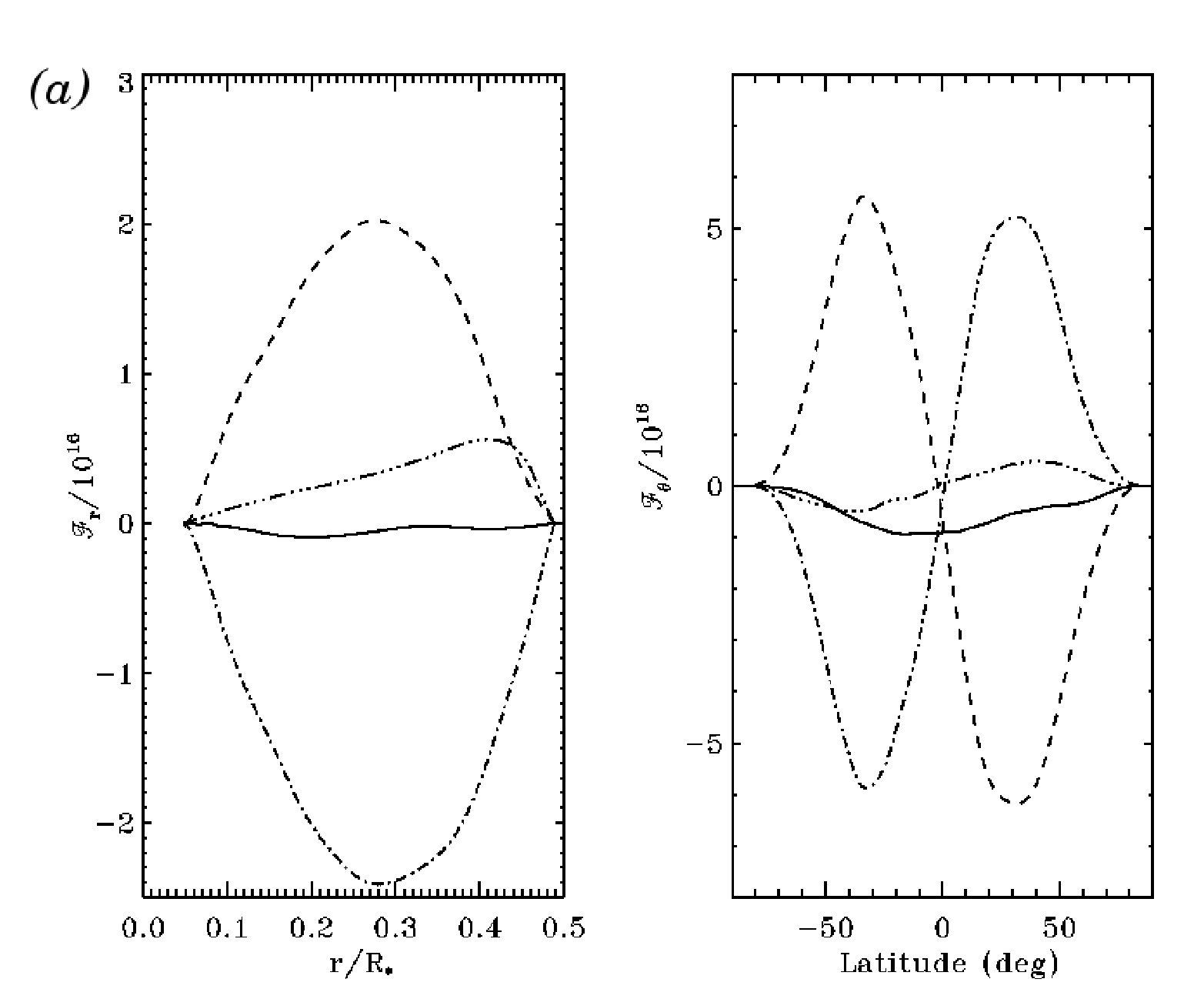}%
\includegraphics[height=5.5cm,width=0.5\textwidth,clip]{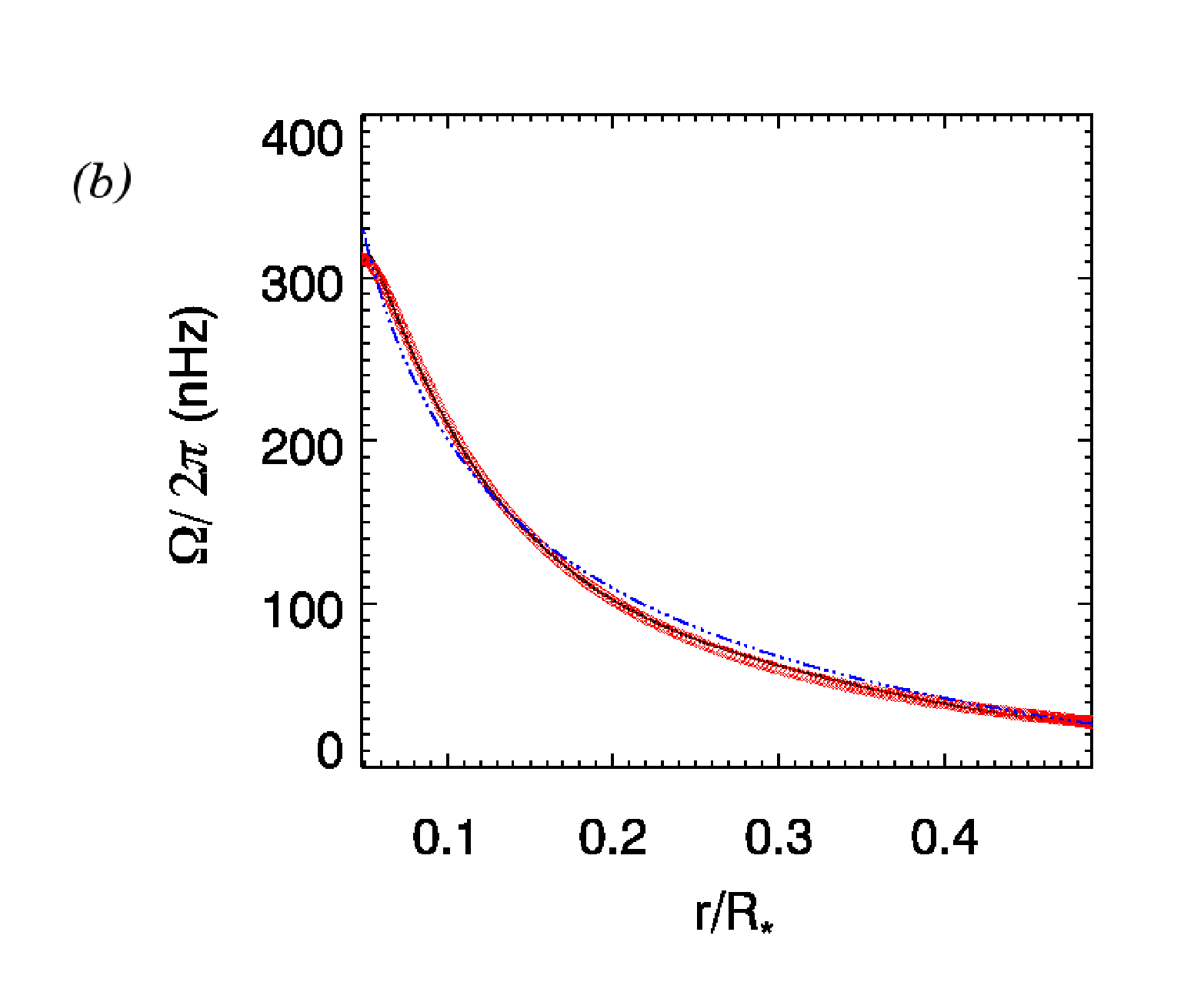}
\caption{$(a)$ Temporal average of the latitudinal line integral of the angular
momentum flux ${\cal F}_r$ (divided by $R^2$), and of the radial line integral of
the angular momentum flux $\cal{F}_\theta$. The fluxes are decomposed into their viscous (dashed-double dotted lines), Reynolds
stress (dot-dashed lines), and meridional circulation (long dashed lines)
components. The solid curves represent the total fluxes and serve to
indicate the quality of the stationarity achieved in the simulation. The fluxes have been averaged over a period of 1800 days.
 $(b)$ Collapsed angular velocity profile along the radial
  direction for a temporal average of 1800 days. The double-dotted dashed
  line represents a 10\% quality fit (see text) and the squares represent a
  better quality (2\% at worse) fit achieved using a ratio of polynoms.}
\label{fig3}
\end{center}
\end{figure}

In Fig~\ref{fig3}.a, we have represented these different contributions to the
radial and latitudinal fluxes by integrating over concentric shells and
cones with different angles respectively, and by subsequently averaging
over the last 1800 days of the simulation. As it can be seen from the solid
lines representing the total flux, the simulation is now balanced, since
this line is near zero. The radial viscous flux is small and positive, this
sign being expected from the angular velocity gradient. The transport by
meridional circulation and Reynolds stresses are of opposite signs and
balance each other in radius and latitude. The meridional circulation
radial flux is positive, which indicates an outward transport of angular
momentum, whereas in latitude it is negative in the northern hemisphere,
corresponding to a poleward transport of angular momentum. \\ In
Fig.~\ref{fig3}.b, we present the collapsed angular velocity profile obtained
by temporal and angular (latitudinal and longitudinal) average of the
$v_\phi$ component of the velocity over more than 10 rotation periods
(e.g. 3230 days).\\ The resulting profile confirms the strong differential
rotation seen at all latitudes in Fig.~\ref{fig2}.a. Although the angular
velocity does not match a $\propto r^{-2}$ relation, which would correspond
to conservation of the specific angular momentum within the computational
domain, it is clearly a decreasing function of the radius, contrary to what
is commonly assumed in 1D stellar evolution models. The profile we obtain
is not described by any simple function, although a 10\% quality fit can be
obtained with a function of the form $$f(r) = a + b r^n$$ with a = -141, b
= 122 and n = -0.45. On the plot in Fig.~\ref{fig3}.b, we have also used a ratio
of two polynomial functions in order to achieve a better fit
(squares over the solid line). The complexity of the function mirrors the
intricate combination of meridional circulation, viscosity and Reynolds
stresses in transporting the angular momentum, and the important role of
Reynolds stresses, which are the main competitor of meridional
circulation. The sensitivity of this rotation law to various parameters
of the simulation is currently under investigation (Brun \& Palacios \cite{BP06}).

\section{Back to 1D modelling}

In this paper, we have presented the first simulation of turbulent rotating
convection in the inner part of the extended convective envelope of a RGB
low-mass star. This first case is moderately turbulent. The convection is
strong, and structured in large cells that cover the top of the
computational domain and become smaller at the base of the domain, in part
due to geometrical effects. The developed convective state realized in our
simulation is associated with very large temperature fluctuations and
radial velocities, orders of magnitude larger than the ones found in
similar simulations of the solar convection zone (Brun \& Toomre \cite{BT02}). The differential
rotation, which is nearly cylindrical, presents large gradients in both the
latitudinal and radial directions. An average over all longitudes and
latitudes reveals a mean radial profile far from the uniformity that has
been generally assumed in 1D stellar evolution codes. For the present
simulation, we obtain an acceptable fit using a much more intricate
function than we would have expected in the case of conservation of the
specific angular momentum.\\ Although we only explored one point in a large
parameter space, which invites us to be cautious when drawing conclusions from
this first run, the rotation law that emerges from the simulation presented
here is much closer to the case of specific angular momentum conservation
than to that of uniform rotation. This is consistent with what was expected
from both observations and 1D models (see \S~1).\\ This encourages us to
push further our investigation in order to be able to use the rotation
profile predicted by the 3D simulation in 1D stellar evolution models
(Palacios et al., in prep.).

\begin{acknowledgments}
A.P. acknowledges post-doctoral fellowship from the Centre d'\'Etudes
Atomiques. Computations have been performed at CEA-CCRT and CNRS-IDRIS
supercomputer centers.
\end{acknowledgments}

\begin{discussion}

\discuss{Hans-G. Ludwig}{In view of the stochastic excitation of pulsations: are
pulsations present in your model ?}

\discuss{A. Palacios}{No. ASH is an Anelastic code, and acoustic modes are filtered.}

\discuss{F. Kupka}{Wouldn't it be necessary to run many more
  different simulations probing a much larger range of parameters before
  you can extract scaling laws that yo use in your 1D stellar evolution models ?} 

\discuss{A. Palacios}{Yes, we're in the process of running further
 simulations, in particular a more turbulent simulation, but these are time
 consuming computations. The important point from these first results is
 that the rotation appears not to be uniform in the inner convective
 envelope of the red giant. This is different from the assumption that
 people normally do in stellar evolution. It is however encouraging because
 it seems to confirm the feeling that we had from observations, in
 particular from observations of the surface rotation velocities of
 horizontal branch stars.}

\discuss{Comment by F. Kupka}{O.K, because from your conclusions I had the
 impression you're already in the process of putting you results back into
 the evolution code. }

\end{discussion}

\end{document}